# Asymptotic expansions for field moments of bound states

G.W. Forbes[1] and Miguel A. Alonso[2,3]

[1] Macquarie University, Department of Physics and Astronomy, North Ryde 2109, NSW, Australia
[2] Aix Marseille Univ, CNRS, Centrale Marseille, Institut Fresnel, Marseille, France
[3] The Institute of Optics, University of Rochester, Rochester NY 14627, U.S.A.

E-mail: forbes@bigpond.net.au and miguel.alonso@fresnel.fr

**Abstract**

Asymptotic expansions are presented for the moments of bound states in one-dimensional anharmonic potentials. The results are derived by using the SAFE method and include only the first non-zero wave-related correction to the familiar semi-classical approximation. Application to a couple of widely studied potentials that do not permit closed-form solutions is used to demonstrate surprising accuracy even in cases that are far from any asymptotic limit. We explore the absence of alternate terms in the asymptotic expansions as a way to explain the accuracy of the end results. Those results are expressed in terms of definite integrals with integrands involving the parameter used in the SAFE method to control the extent of the associated elemental field contributions. Importantly, the integrals themselves are shown to be precisely independent of that parameter. Further, although the derivation proceeds by way of an asymptotic expansion for the wavefield that involves the associated classical motion, those entities do not appear in the end results which are expressed in terms of just the potential function and its first four derivatives.

## 1. Introduction

It is appropriate in this special issue to start with a reference to the seminal report from 1972 by Berry and Mount entitled "Semiclassical approximations in wave mechanics"[1]. Even though it appeared 50 years ago, the authors state that "Semiclassical methods are as old as quantum theory itself, and the literature is correspondingly enormous." A specific topic of importance is the wave mechanics of a particle in a potential well. Since exact solutions are known for only a modest collection of potential functions, asymptotic treatments have continued to receive extensive attention and development. Although progress continues[2], asymptotic expansions for the associated eigenvalues are relatively well understood. Uniform approximations for their wavefunctions, on the other hand, have proven to be more challenging.

When applied to wave problems, asymptotic methods are valuable not only because they usefully approximate solutions, but also because they provide a link to more basic and intuitive (if incomplete) physical models such as classical mechanics and the ray model for optical or mechanical waves. The simplest approaches, however, encounter problems in the form of localised divergent errors near turning points when applied to finding modes/eigenstates, or at caustics in the more general case. Alternative approaches have been proposed to avoid these issues, some based on Fourier representations[3,4] (which remove the divergences at the cost of introducing analogous problems in momentum space), or on more uniform asymptotic approximations[5].

Yet another strategy for avoiding such errors is to express the wavefunction as a superposition of localised waves, whose profiles are often chosen to be Gaussian. To our knowledge, this approach was first used in seismology[6], and then proposed independently for quantum mechanics[7]. More than twenty years ago we proposed a related asymptotic method, founded on intuitive approaches to working with semi-classical analysis[8-12]. Our initial focus was on applications such as ray and wave optics as well as classical and quantum mechanics. This meant we emphasised a variety of familiar linear partial differential equations (PDEs). The definitive characteristic of the method was the step of expressing a wavefield as a superposition of elemental fields that (a) are not required to individually satisfy the wave equation of interest and (b) are able to be reconfigured by modifying the value of a parameter, written as $\gamma$, that controls their widths. For the net result to consistently yield the desired solution, the superposition must be asymptotically independent of that reconfiguration. We referred to the result as a stable aggregate of flexible elements (SAFE). By construction, the SAFE method aims to provide a mathematical framework that formalises the mental flexibility that physicists long ago learned to exercise when modelling phenomena with the likes of ray and wave theories: its reconfiguration effectively bridges the notions of a sum of ray- or particle-like contributions to a sum of wave-like components. It turned out that the method created sound and accurate asymptotic solutions.

In this work we focus on the bound states of one-dimensional anharmonic oscillators, i.e. potential wells with non-parabolic profiles. Varieties of such potential wells have been studied extensively and serve as simple models for a range of physical systems. The underlying equation of interest in one dimension can be expressed as

$$U''(x) + k^2[\varepsilon - V(x)]U(x) = 0, \tag{1}$$

where asymptotic methods serve to approximate solutions for the eigenstate $U$ and the eigenvalue $\varepsilon$ as the asymptotic parameter $k$ becomes large. (In quantum-mechanical contexts, $k$ is inversely proportional to Planck's constant.) Of special interest here are cases where the potential $V$ is smoothly varying and has a local minimum, and there is a set of discrete values of $\varepsilon$ which admit solutions that are localised to be near the well within the region defined by $V(x) < \varepsilon$. Closed-form solutions exist for these eigenvalues and bound states for only a modest number of simple potential functions. More generally, the solutions can be explored only through approximate schemes such as numerical or asymptotic methods.

Particular aspects of interest in such applications are the moments of those bound states, defined as

$$\mathcal{M}_m = \int x^m |U(x)|^2 \mathrm{d}x. \tag{2}$$

This work focuses precisely on asymptotic expansions for these moments of the form

$$\mathcal{M}_m = \sum_{j=0}^{\infty} \frac{1}{k^j} \mathcal{M}_{mj}. \tag{3}$$

It turns out that these asymptotic coefficients are ultimately able to be expressed in terms of single integrals of slowly-varying expressions involving the potential and its derivatives, as well as the estimated eigenvalues. That is, *our goal is to derive asymptotic estimates for the moments that do not require the explicit asymptotic construction of the bound state wavefunctions*. These results are derived by using the SAFE method, and hence involve the parameter $\gamma$ that sets the width of the contribution for each ray/classical path. Surprisingly, while the integrands in the end results include $\gamma$, the resulting integrals are shown to be independent of this parameter. The value of $\gamma$ can therefore be chosen to facilitate any required numerical integration. We know of no other way to derive directly such explicit asymptotic expansions of the moments of the bound states.

The main results of this work, namely the semiclassical estimates for the moments, are reported in Section 3. Before that, though, Section 2 presents a brief treatment of standard approaches as well as an extension of the asymptotic construction of the wavefunctions and eigenvalues by using the SAFE method. Some illustrative applications are also presented in each section.

## 2. Review of Asymptotic Solutions

To give insight related to the main results of Section 3, we start with a review of the standard asymptotic solution to Eq.(1). We also present a necessary extension of the SAFE solution and use the result to show how this method can itself be used to derive an asymptotic expansion of the eigenvalues.

### 2.1 The JWKB approach

The conventional step for constructing the two linearly independent asymptotic solutions is to express the field as an exponential, say

$$U(x) = u_0\, \mathrm{e}^{\mathrm{i}\, k\, S(x)}, \tag{4}$$

where $u_0$ is constant. With that, Eq.(1) becomes

$$\varepsilon - V(x) - S'^2(x) - \frac{S''(x)}{\mathrm{i}\, k} = 0. \tag{5}$$

When considering the behaviour of the solutions for large $k$, it is effective to expand $S(x)$ as a series in the form

$$S(x) = \sum_{n=0}^{\infty} S_n(x)/(\mathrm{i}k)^n. \tag{6}$$

Upon substituting this ansatz into Eq.(5), the coefficient of each power of $k$ can then be required to vanish independently. With this, the topmost few orders that emerge are

$$S_0'(x) = \pm\sqrt{\varepsilon - V(x)}, \tag{7}$$

$$S_1'(x) = -\frac{S_0''(x)}{2S_0'(x)}, \tag{8}$$

$$S_2'(x) = -\frac{S_1'^2(x) + S_1''(x)}{2S_0'(x)}. \tag{9}$$

It follows from Eqs.(7) and (8) that —to within an irrelevant constant that can be absorbed within $u_0$— $S_1(x)$ is just $\frac{-1}{4}\log[\varepsilon - V(x)]$ and this then yields the two familiar approximate solutions that can be written as:

$$U_\pm(x) = u_0 \frac{1 + \frac{1}{ik}S_2^\pm(x) + O(k^{-2})}{[\varepsilon - V(x)]^{1/4}} e^{\pm i k \int^x \sqrt{\varepsilon - V(s)}\, ds}. \tag{10}$$

In this way, not only is a second-order ODE reduced to a series of simple first-order ODEs, but the end result is expressed in terms of integrals of only slowly varying functions. The main drawback is that $U_\pm(x)$ diverges at any classical turning point, i.e. wherever $\varepsilon - V(x)$ vanishes; a true solution to Eq.(1) simply has an inflection point there. Importantly, Eq.(10) reveals that in regions where $\varepsilon - V(x)$ is positive (or negative), the solutions are rapidly oscillating (or decaying/growing, respectively) once $k$ is large.

The solution in Eq.(10) is often referred to as the WKB or JWKB approximation [13]. Notice that the sign chosen in Eq.(7) impacts $S_2'(x)$, but has no impact on $S_1'(x)$. That is why $S_2(x)$ has been given a superscript of $\pm$ in Eq.(10). More generally, that sign does not impact $S_n(x)$ for odd values of $n$, but sets the global sign on $S_n'(x)$ for even values. Unlike $S_1(x)$, although the solution for $S_2^\pm(x)$ follows from Eqs.(5-7), it cannot be found in closed form. Curiously, while $S_n(x)$ can be found in closed form for all odd values of $n$, the even cases all involve unresolved indefinite integrals.

An insight into this asymptotic solution follows upon expressing $U(x)$ as

$$U(x) = A(x)\, e^{ik\phi(x)}, \tag{11}$$

in which case, in place of Eq.(5), the following is now required to vanish

$$\varepsilon - V(x) - \phi'^2(x) - \frac{2A'(x)\phi'(x) + A(x)\phi''(x)}{ikA(x)} + \frac{A''(x)}{k^2 A(x)}, \tag{12}$$

Making the coefficient of $O(k^{-1})$ vanish leads to a first-order ODE for $A(x)$ that can be solved exactly:

$$A(x) = u_0/\sqrt{\phi'(x)}. \tag{13}$$

Note that the constant $u_0$ can absorb the sign ambiguity associated with the square root in Eq.(13). With this, the condition associated with Eq.(12) is reduced to

$$\varepsilon - V(x) - \phi'^2(x) + \frac{3\phi''^2(x) - 2\phi'(x)\phi^{(3)}(x)}{4k^2\phi'^2(x)} = 0, \tag{14}$$

hence only even powers of $k^{-1}$ are needed in an asymptotic solution for $\phi(x)$. With

$$\phi(x) = \sum_{n=0}^\infty \phi_{2n}(x)/(ik)^{2n}, \tag{15}$$

the two linearly independent solutions can now be expressed as

$$U_\pm(x) = \frac{u_0}{\sqrt{\phi'(x)}} e^{\pm i k\, \phi(x)}, \tag{16}$$

where the first couple of terms are found simply by using Eq.(14):

$$\phi_0'(x) = \sqrt{\varepsilon - V(x)}, \tag{17}$$

$$\phi_2'(x) = \frac{2\phi_0'(x)\phi_0^{(3)}(x) - 3\phi_0''^2(x)}{8\phi_0'^3(x)}. \tag{18}$$

Going to higher orders readily yields explicit expressions for $\phi_{2n}'(x)$ for $n > 1$. Notice that the constants of integration from the likes of Eqs.(17) and (18) simply modify $u_0$ and that the solution's amplitude, namely $u_0/\sqrt{\phi'(x)}$, is then given explicitly in terms of the associated integrands.

It is helpful to note that the relation between the two asymptotic solutions in Eqs.(4) [via (10)] and (16) follows upon expressing the second version as

$$U_\pm(x) = \frac{u_0\, e^{\pm i k \phi(x)}}{\sqrt{\phi'(x)}} = u_0\, e^{ik\{\pm\phi(x)-\log[\phi'(x)]/(2ik)\}}. \tag{19}$$

Upon introducing Eq.(15) and comparing Eq.(19) with Eq.(4), it can be seen that the even-order coefficients of Eq.(6) satisfy

$$S_{2n}(x) = \phi_{2n}(x). \tag{20}$$

The odd-order coefficients, on the other hand, follow upon asymptotically expanding the logarithm in Eq.(19) and using Eq.(20) to see directly that

$$S_3(x) = -\frac{\phi_2'(x)}{2\phi_0'(x)} = -\frac{S_2'(x)}{2S_0'(x)}, \tag{21}$$

$$S_5(x) = \frac{S_2'^{\,2}(x) - 2S_0'(x)S_4'(x)}{4\,S_0'^{\,2}(x)}. \tag{22}$$

As indicated in Eq.(21), in view of Eq.(20), either $\phi$ or $S$ can appear throughout the right-hand side of these equations. This clarifies the different character of alternate terms in Eq.(6). The key is that, as shown in Eqs.(21) and (22), the odd coefficients in the expansion for $S(x)$ can be found explicitly in terms of the first derivative of lower-order (even) terms and these are given explicitly by the likes of Eqs.(17) and (18).

It follows that every other integral encountered when working through the original series of Eqs.(7), (8), (9), etc. can be avoided. Notice also that while the $2n+1$ terms $S_j(x)$ for $j = 0,1,2 \ldots 2n$ yield the conventional solution via Eqs.(4) and (6) to within an error of $O(k^{-2n})$, just the even members of that set —the $n+1$ terms $S_{2j}(x) = \phi_{2j}(x)$ for $j = 0,1,2 \ldots n$— can give the solution via Eqs.(15) and (16) to within an error of $O(k^{-(2n+1)})$. The reduced number of integrals and the greater accuracy in the latter case is because the odd terms essentially come for free when handled in this way. When it comes to field moments, it is worth noting that the asymptotic expansion of the field amplitude given in Eqs.(13) and (15) involves only even powers of $k^{-1}$.

*2.2 The SAFE method*

Here we provide a brief review of the basic elements of the SAFE method presented in Refs. 8 to 12. For brevity, we adopt a convention for citing equations from Refs. 8, 9, and 10, whose titles included the roman numerals I, II, and III, respectively, by preceding their equation number with either I-, II-, or III-. For example, the generic SAFE decomposition of a field is based on Eq.(I-2.15) and written here as

$$U(x) = \sqrt{\frac{k}{2\pi}} \int A(\tau,\gamma)\sqrt{Y'(\tau)}\, g(x,\gamma,\tau)\, d\tau, \tag{23}$$

where

$$g(x,\gamma,\tau) = e^{k\left(-\frac{\gamma}{2}[x-X(\tau)]^2 + i\{L(\tau)+[x-X(\tau)]P(\tau)\}\right)}, \tag{24}$$

with

$$Y(\tau) = \gamma\, X(\tau) + i\, P(\tau). \tag{25}$$

That is, the field is expressed as a sum of components parametrized by $\tau$ where each has a Gaussian envelope with a width determined by $k\gamma$. (Note that the standard asymptotic ansatz in Eq.(11) corresponds to $\gamma \to \infty$, while $\gamma \to 0$ leads to an analogous momentum-based approximation.) The individual Gaussians are centred at $X(\tau)$ and modulated with a frequency $kP(\tau)$. Their complex weight is given by $A(\tau,\gamma)\sqrt{Y'(\tau)}\, e^{ikL(\tau)}$ which was configured to force the integral in Eq.(23) to be asymptotically independent of $\gamma$. That insensitivity is ensured by requiring

$$L'(\tau) = P(\tau) X'(\tau). \tag{26}$$

Although the complexity of Eqs.(23) and (24) may seem overwhelming at first sight, the elements become intuitive upon considering the context of mechanics. When Eq.(23) is used in Eq.(1) and the steps of integration by parts are applied as in Eqs.(I-4.2) and (I-4.8), it follows that $X(\tau)$ and $P(\tau)$ can be identified as the classical position and momentum while $L(\tau)$ is the classical action. This means that the parametric curve described by $\{X(\tau), P(\tau)\}$ is the familiar phase space curve (PSC) or Lagrange manifold. It is convenient to choose $X'(\tau)$ to be proportional to $P(\tau)$, say

$$X'(\tau) = \chi P(\tau), \tag{27}$$

(hence if $\tau$ represents time, $\chi$ is effectively the inverse of the mass). In this case, when $A(\tau, \gamma)$ is expanded as an asymptotic series, say

$$A(\tau, \gamma) = \sum_{n=0}^{\infty} A_n(\tau, \gamma)/(ik)^n, \tag{28}$$

it turns out that $A_0(\tau, \gamma)$ is a constant, say $A_0(\tau, \gamma) = a_0$.

Upon substituting Eq.(23) into Eq.(1), integrating by parts, collecting terms and making the different orders vanish as for Eq.(I-4.11), it is found that

$$P^2(\tau) = \varepsilon - V[X(\tau)], \tag{29}$$

which together with Eq.(27) yield both $X(\tau)$ and $P(\tau)$. It is also found that

$$A_1(\tau, \gamma) = a_0 \left\{ F_1(\tau) - \frac{i}{48} \left[ 5 \left( \frac{1}{X'(\tau)Y'(\tau)} \right)'' - \frac{X'(\tau)Y'''(\tau) + X'''(\tau)Y'(\tau)}{[X'(\tau)Y'(\tau)]^2} \right] \right\}, \tag{30}$$

where

$$F_1'(\tau) = \frac{-5\chi^5 V'^2[X(\tau)] - 4\chi^3 X'^2(\tau) V''[X(\tau)]}{32 X'^4(\tau)}. \tag{31}$$

In contrast, it turns out that $A_2(\tau, \gamma)$ can be found in closed form and is given by an expression of the form

$$A_2(\tau, \gamma) = \frac{1}{2a_0} A_1^2(\tau, \gamma) + a_0 \, \mathcal{F}(\tau), \tag{32}$$

where

$$\mathcal{F}(\tau) = f_2 + \frac{\chi^2 (2\gamma^2 - V'')^2 (5Y''^2 - 2Y'Y^{(3)})}{64 Y'^6} + \frac{\chi^3 (2\gamma^2 - V'')(\chi V'Y' + 5PY'') - i\chi(9Y''^2 - 4Y'Y^{(3)})}{96 Y'^5} V^{(3)}$$

$$-\chi^2 \frac{2Y'^2 + (\gamma\chi Y' - 7iY'')P}{96 Y'^4} V^{(4)} - \frac{i\chi X'^2}{96 Y'^3} V^{(5)}. \tag{33}$$

where the argument $\tau$ of $P, X, Y$ and their derivatives is suppressed for brevity, and $V$ and its derivatives are evaluated at $X(\tau)$. This second correction is a new result that was not reported in Refs. 8-12. Note that Eqs.(30) to (33) bring Eq.(6) to mind where alternate coefficients can be found in closed form while the others involve unresolved indefinite integrals. Importantly, the singularities in each piece of Eq.(30) whenever $X'(\tau)$ vanishes ultimately cancel out so that $A_1(\tau, \gamma)$ is singularity-free, as is $\mathcal{F}(\tau)$.

## 2.3 SAFE-based estimation of the eigenvalues

For the bound states investigated below, the PSC is a closed loop and both $X(\tau)$ and $P(\tau)$ are periodic functions. Much like the case in Eq.(II-5.1), the integral in Eq.(23) is then taken to span a single period. For that integrand to be periodic, its phase must change by a multiple of $2\pi$ with each circuit of the PSC. Note from Eq.(26) that the difference in $L$ for two values of $\tau$ corresponds to the area under the PSC between the two corresponding points. If the parametrization starts at the rightmost point on the PSC (where $X(\tau)$ is a maximum and $P(\tau)$ is zero) and proceeds in a counterclockwise manner, $L(\tau)$ is then a decreasing function that will, after completing the loop, decrease by an amount equal to the phase space area enclosed by the PSC. Note also from Eq.(25), assuming that $\gamma$ is real and positive, that $Y$ maps the phase space onto a complex plane, where the position axis is scaled by $\gamma$. This means that the phase of $Y'(\tau)$ equals the angle from the horizontal to the local tangent to the corresponding point of this scaled PSC. Therefore, the phase of the factor $\sqrt{Y'(\tau)}$ inside the integrand of Eq.(23) increases by $\pi$ during a circuit. At zeroth order therefore, $k\,L(\tau)$ must fall by an odd multiple of $\pi$ with each circuit, say $(2n+1)\pi$, and, by using Eqs.(26) and (29), this yields a zeroth-order approximation for the eigenvalue, say $\varepsilon \approx \varepsilon_n^{(0)}$ for Eq.(1) and effectively discretises the permissable areas for the PSC loop.

Things are more interesting at higher orders. To begin, notice that all but $F_1(\tau)$ is periodic on the right-hand side of Eq.(30) so it may appear to be the only source of a change in $A_1(\tau,\gamma)$ with each circuit of the PSC. It turns out that $\text{Im}[A_1(\tau,\gamma)]$ is not only periodic, but it can be found in a singularity-free closed form. This means that the change in $A_1(\tau,\gamma)$, written here as $\delta F$, is real and is determined after cancelling the singularities in the separate pieces of Eq.(30). Depending on the form of $V(x)$, $\delta F$ need not be zero.

Importantly, a higher order asymptotic correction emerges for the eigenvalues when $\delta F \neq 0$. To see this, the first step follows upon observing that $A(\tau,\gamma)$ of Eq.(28) can be re-expressed as

$$a_0 + \frac{A_1}{ik} + \frac{A_2}{(ik)^2} + O(k^{-3}) = \left[a_0 - \frac{1}{k^2}\left(A_2 - \frac{A_1^2}{2a_0}\right) + O(k^{-4})\right] e^{-ik\left[\frac{A_1}{a_0 k^2} + O(k^{-4})\right]}. \tag{34}$$

where the arguments of $A_n(\tau,\gamma)$ have been suppressed. This result is easily extended to any order by using an inverse of the process used in Eq.(19), i.e. first expand the logarithm of the expression on the left-hand side of Eq.(34) and then re-expand the exponential of only the even components. Notice that the end result involves two series that hold only even powers. This is evidently an alternative structure for the SAFE method (where, reminiscent of the alternative JWKB solution of Section 2, both $A$ and $L$ are expanded as even-power-only asymptotic series). It follows from Eqs.(23), (24), and (34), that the refined quantization condition is that $k\,L(\tau)$ change by $-(2n+1)\pi + \delta F/k$. In this way, SAFE leads to the first asymptotic correction for the eigenvalue, say $\varepsilon \approx \varepsilon_n^{(1)}$.

In keeping with well known results, the error in $\varepsilon_n^{(1)}$ is of $O(k^{-4})$. This follows here from two observations: first, when $A_2(\tau,\gamma)$ is included, it is $A_2 - \frac{A_1^2}{2a_0}$ that enters Eq.(34) and second, that $\mathcal{F}$ of Eq.(32) is periodic. Taken together, these mean that the factor in front of the exponential in Eq.(34) is periodic hence $A_2(\tau,\gamma)$ brings no refinement of the quantization condition; if there is a change in $A_3(\tau,\gamma)$ with each circuit, that would ultimately sit within the exponential of Eq.(34) and lead to a correction at $O(k^{-4})$. We offer these comments to show how SAFE also leads to these results.

*2.4 Specific potentials for demonstration*

In this section we illustrate the accuracy of the asymptotic eigenvalue expansions. This sets a useful reference point for the moment-related results that follow.

The well known Morse potential[14] can be written as

$$V(x) = e^{-2x} - 2e^{-x}, \tag{35}$$

and serves as a simple testing ground for these ideas. The associated eigenvalues and bound states are known in closed form along with $X(\tau)$, $P(\tau)$, and $L(\tau)$, i.e. the classical position, momentum, and action. Further, it turns out that $\varepsilon_n^{(0)}$ is equal to the exact eigenvalue and $A_1(\tau,\gamma)$ can also be found in closed form (and seen to be periodic, hence $\delta F = 0$). That is, SAFE's asymptotic approximations are all easily accessible in this case. Because the exact results are known, however, this case can only serve as a sanity check.

On the other hand, the Pöschl-Teller potential[15] given by

$$V(x) = -\text{sech}^2 x, \tag{36}$$

has associated bound states for Eq.(1) that are known in closed form only for the special cases where $k^2 = \kappa(\kappa+1)$ for some positive integer $\kappa$. Non-integral values of $\kappa$ provide valuable cases for demonstration in what follows. More generally, it turns out that the discrete eigenvalues are known to be

$$\varepsilon_n = -\left(\sqrt{1+\frac{1}{4k^2}} - \frac{2n+1}{2k}\right)^2 = \frac{-(\kappa-n)^2}{\kappa(\kappa+1)}, \tag{37}$$

for $n = 0,1,2\ldots\lfloor\kappa\rfloor$, hence $-1 < \varepsilon_n < 0$. When there is no closed form for the bound states (i.e. $\kappa$ is nonintegral) the likes of SAFE's asymptotic approximations become valuable.

Together, Eqs.(26), (27), and (29) lead in this case to

$$X(\tau) = \text{arcsinh}\left(\sqrt{\frac{1+\varepsilon}{-\varepsilon}} \cos \tau\right), \quad (38)$$

$$P(\tau) = \sqrt{-\varepsilon(1+\varepsilon)} \frac{\sin \tau}{\sqrt{1-(1+\varepsilon)\sin^2 \tau}}, \quad (39)$$

$$L(\tau) = \sqrt{-\varepsilon}\,\tau - \arctan(\sqrt{-\varepsilon}\tan\tau) - \pi\left\lfloor\frac{\tau}{\pi} + \frac{1}{2}\right\rfloor. \quad (40)$$

It follows from Eq.(40) that the change in $k\,L(\tau)$ with each circuit of the PSC is $-k2\pi(1-\sqrt{-\varepsilon})$, hence as discussed in the paragraph following Eq.(32),

$$\varepsilon_n^{(0)} = -\left(1 - \frac{2n+1}{2k}\right)^2. \quad (41)$$

By using Eqs.(30) and (31), $A_1(\tau,\gamma)$ can also be found in closed form and it turns out that $\delta F = \pi/4$, which leads to an asymptotic correction for the eigenvalue:

$$\varepsilon_n^{(1)} = -\left(1 - \frac{2n+1}{2k} + \frac{1}{8k^2}\right)^2. \quad (42)$$

As discussed at the end of Section 2.3, Eqs.(41) and (42) are accurate to within terms of $O(k^{-2})$ and $O(k^{-4})$, respectively, as can be confirmed in this case by reference to Eq.(37).

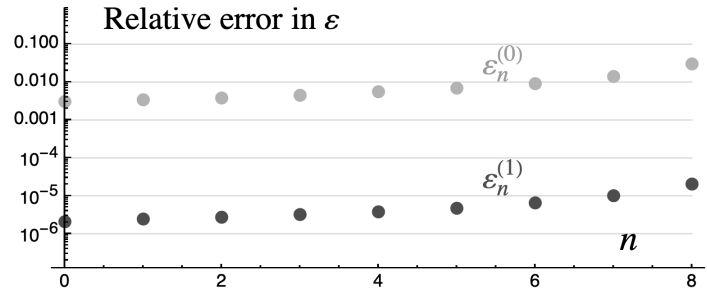

**Figure 1** Plot of relative errors in asymptotic estimates of eigenvalues of Eqs(41) and (42) for $\kappa = 8.9$.

As shown in Fig.1, even in the seemingly not very asymptotic case of $\kappa = 8.9$ where there are only nine bound states, it is found that $\varepsilon_n^{(1)}$ is typically accurate to better than 10 parts per million. These eigenvalues are overlaid on a plot of the potential in Fig.2. Note that although the relative errors are increasing with $n$, that increase does not continue because Fig.1 includes all of the bound states; while there are more bound states for higher values of $\kappa$, the associated error plots have a similar shape to those in Fig.1 with relative errors of reduced magnitude. The associated PSCs as well as the nine bound states are plotted in Fig.3. These bound states have been determined by numerically integrating Eq.(1).

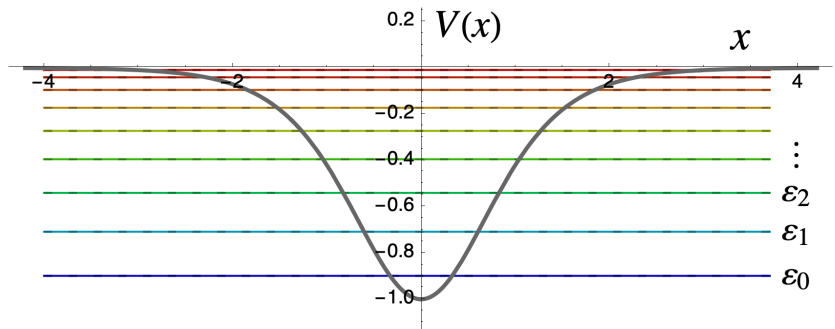

**Figure 2** Plot of the potential of Eq.(36) together with the eigenvalues of Eq.(37) with $\kappa = 8.9$ which are indistinguisable from the overlaid dashed approximation given in Eq.(42).

While it is not the focus of this work, we note that SAFE estimates for the bound states can be readily determined by using numerical quadrature on Eq.(23) and achieve accuracies that typically start well below 10% of peak to valley at zeroth order, i.e. with $A(\tau,\gamma) = a_0$, and see steady improvement as the corrections $A_1(\tau,\gamma)$ and $A_2(\tau,\gamma)$ are included. This means that, when those two corrections are included, plots of the SAFE estimates for the bound states are effectively indistinguishable from the numerical results presented in Fig.3. It is worth noting too that the selection of optimal values for $\gamma$ to maximize insensitivity are discussed at Eq.(I-5.11). It is important to note, however, that the novel result in Eqs.(32) and (33) is required for the key contributions that are presented in the next section.

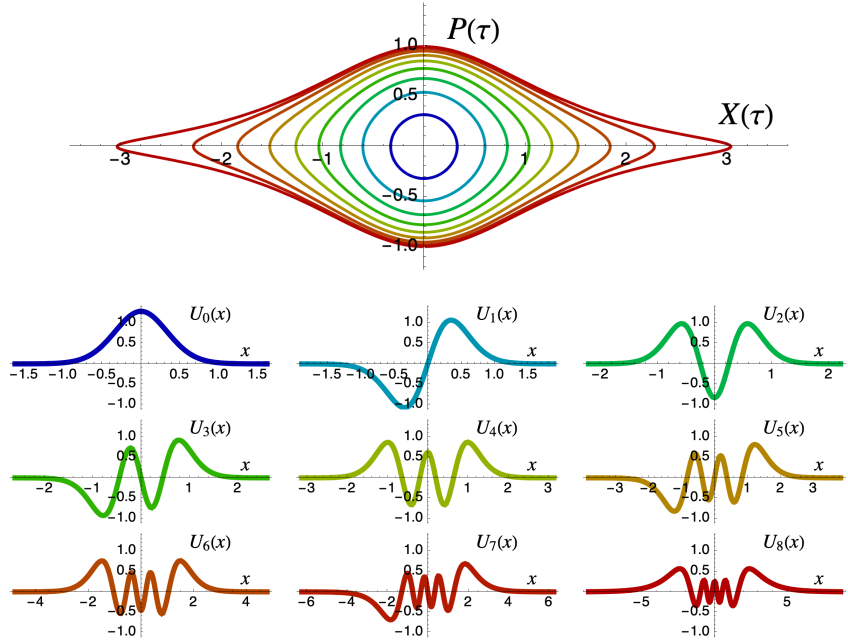

**Figure 3** The PSCs for the bound states of Fig.2 along with the associated normalised bound states in matching colours. Note the changing horizontal axis scale in the lower nine plots.

## 3. Field Moments

We now focus on the main goal of this work: to provide direct asymptotic estimates of the moments of the bound states.

### 3.1 Derivation of the estimates

The derivation of these expressions follows from a simple generalisation of Eq.(III-3.9). The first step involves substituting the SAFE ansatz of Eq.(23) into Eq.(2):

$$\mathcal{M}_m = \frac{k}{2\pi}\int \left[\int A(\xi,\gamma)\sqrt{Y'(\xi)}\, g(x,\gamma,\xi)\, d\xi\right]\left[\int A(\zeta,\gamma)\sqrt{Y'(\zeta)}\, g(x,\gamma,\zeta)\, d\zeta\right]^{*} x^m dx, \qquad (43)$$

where the asterisk denotes complex conjugation. Because $g(x,\gamma,\tau)$ of Eq.(24) depends on $x$ only through the exponential of a quadratic, the integral over $x$ in Eq.(43) can be evaluated first in closed form for any integer $m$. That leaves behind the integrals over $\xi$ and $\zeta$ which can be reduced asymptotically to a single integral by using the saddle-point method as discussed at Eq.(II-A4). Upon first changing the variables of integration to $\tau = \xi + \zeta$ and $\delta = \xi - \zeta$, it is readily seen that the exponent of the integrand is stationary when $\delta = 0$. Application of the saddle-point method then leaves just the integral over $\tau$ of an asymptotic series.

As a demonstration, consider low-order approximations for the first few moments, say $\mathcal{M}_{mj}$ of Eq.(3) for $m$ and $j$ both ranging from 0 to 2. The process described above leads to results that involve $\mathrm{Im}[A_1(\tau,\gamma)]$ and $\mathrm{Re}\left\{A_2(\tau,\gamma) - \frac{1}{2a_0}[A_1(\tau,\gamma)]^2\right\}$, which both follow from Eqs.(30) and (32). Note that the solution to Eq.(31) leaves a constant of integration in $A_1(\tau,\gamma)$. That constant is written as $f_1$. Similarly, the $\mathcal{F}(\tau)$ that appears in $A_2(\tau,\gamma)$ carries an analogous arbitrary constant denoted by $f_2$ in Eq.(33).

The end result for the zeroth moment can be written as

$$\mathcal{M}_0 = \int a_0^2 \, d\tau + \frac{2a_0^2}{k} \int \mathrm{Im}(f_1) \, d\tau + \frac{a_0^2}{k^2} \int \{\varphi + \mathcal{K}_0[X(\tau), Q(\tau)]\} \, d\tau + O(k^{-3}), \tag{44}$$

where $Q(\tau) = V'^2[X(\tau)] + [2\gamma X'(\tau)/\chi]^2$, and $\mathcal{K}_0(x,q)$ and the constant $\varphi$ are given by

$$\varphi = 2\{[\mathrm{Im}(f_1)]^2 - \mathrm{Re}(f_2)\}, \tag{45}$$

$$\mathcal{K}_0(x,q) = -\frac{V_4}{12q} - V_1 V_3 \frac{\gamma^2 - 2V_2}{6q^2} - V_2 \frac{(2\gamma^2 - V_2)(6\gamma^2 + V_2)}{12q^2} + V_1^2 V_2 \frac{4\gamma^4 - V_2^2}{3q^3}. \tag{46}$$

Here, $V^{(n)}(x)$ is written as $V_n$. Similarly, the approximation for the first moment is

$$\mathcal{M}_1 = a_0^2 \int X(\tau) \, d\tau + \frac{2a_0^2}{k} \int \mathrm{Im}(f_1) X(\tau) \, d\tau + \frac{a_0^2}{k^2} \int \{\varphi \, X(\tau) + \mathcal{K}_1[X(\tau), Q(\tau)]\} \, d\tau + O(k^{-3}), \tag{47}$$

where $\mathcal{K}_1(x,q)$ is given by

$$\mathcal{K}_1(x,q) = x \, \mathcal{K}_0 - \frac{V_3}{6q} - V_1 V_2 \frac{2\gamma^2 - 3V_2}{12q^2}. \tag{48}$$

It now follows from Eqs.(44) and (47) that the normalised first moment (i.e. the centre of mass) satisfies

$$\frac{\mathcal{M}_1}{\mathcal{M}_0} = \langle X \rangle + \frac{\langle \mathcal{K}_1 \rangle - \langle X \rangle \langle \mathcal{K}_0 \rangle}{k^2} + O(k^{-3}), \tag{49}$$

where

$$\langle f \rangle = \int f(\tau) \, d\tau / \int 1 \, d\tau. \tag{50}$$

Similarly, the normalised second moment is found to satisfy

$$\frac{\mathcal{M}_2}{\mathcal{M}_0} = \langle X^2 \rangle + \frac{\langle \mathcal{K}_2 \rangle - \langle X^2 \rangle \langle \mathcal{K}_0 \rangle}{k^2} + O(k^{-3}), \tag{51}$$

where

$$\mathcal{K}_2(x,q) = 2x \, \mathcal{K}_1 - x^2 \, \mathcal{K}_0 - \frac{\gamma^2}{3q} + V_1^2 \frac{2\gamma^2 - V_2}{3q^2}. \tag{52}$$

Notice that the arbitrary constants of integration, namely $f_1$ and $f_2$, have dropped from Eqs.(49) and (51) and there are no longer any terms of $O(k^{-1})$. Their zeroth-order terms are the familiar semi-classical results, but the terms of $O(k^{-2})$ are wave-related corrections.

There is a useful simplification for the integrals in Eqs.(49) and (51). To see this, notice that the PSC extends between two roots of $\varepsilon - V(x)$, at say $x = x_1$ and $x = x_2$. By using Eqs.(27) and (29), it is possible to change the variable of integration from $\tau$ to $x = X(\tau)$ so that the $\langle \mathcal{K}_n \rangle$ are reduced in this way to a simpler integral expresssed in terms of just the potential function and its derivatives:

$$\langle \mathcal{K}_n \rangle = \int_{x_1}^{x_2} \frac{\mathcal{K}_n[x, \bar{Q}(x)]}{\sqrt{\varepsilon - V(x)}} \, dx \Big/ \int_{x_1}^{x_2} \frac{dx}{\sqrt{\varepsilon - V(x)}}, \tag{53}$$

where $\bar{Q}(x) = V'^2(x) + 4\gamma^2[\varepsilon - V(x)]$. These results can be accompanied naturally by

$$\langle X^m \rangle = \int_{x_1}^{x_2} \frac{x^m}{\sqrt{\varepsilon - V(x)}} \, dx \Big/ \int_{x_1}^{x_2} \frac{dx}{\sqrt{\varepsilon - V(x)}}. \tag{54}$$

A striking characteristic of these results is that, although all $\mathcal{K}_n$ manifestly depend on $\gamma$, it is their averages that enter Eqs.(49) and (51). Because the end results cannot have any dependence on $\gamma$, these averages must be precisely independent of $\gamma$ for all $0 < \gamma < \infty$. It is possible to establish this by, in the case of $\langle \mathcal{K}_0 \rangle$ for example, differentiating the integrand in the numerator of Eq.(53) with respect to $\gamma$ and noting that the indefinite integral over $x$ can then be found in closed form and has a global factor of $\sqrt{\varepsilon - V(x)}$ which vanishes at the endpoints. It follows that $\langle \mathcal{K}_0 \rangle$ is independent of $\gamma$. The same process can be carried out for $\langle \mathcal{K}_1 \rangle$ and $\langle \mathcal{K}_2 \rangle$. Note that, while these integrals are independent of $\gamma$, the value of this parameter can be chosen to make the integrand more slowly varying to facilitate numerical integration. One criterion for such choice is to make $\bar{Q}(x)$ as uniform as possible. This criterion turns out to be consistent with that discussed in Ref. 10 for making the wavefield estimates insensitive to $\gamma$: a scaled PSC given by $\{k^{1/2}\gamma^{1/2}X(\tau), k^{1/2}\gamma^{-1/2}P(\tau)\}$ must not present curvatures of the order of unity or larger. In particular, note that the limiting values of 0 and $\infty$ for $\gamma$ introduce non-integrable singularities in $\mathcal{K}_m$. Recall that these two limits correspond, respectively, to standard JWKB asymptotic approaches in the position and momentum representations. It appears that the asymptotic expansion of the moments requires an "intermediate" representation with finite, nonzero $\gamma$.

It may also be worth observing that, when $V(x)$ is analytic, contour-integration methods can be applied to the integral in the numerator of Eq.(53) where the initial contour connects two square-root branch points on the real axis. Interestingly, the integrand has third-order poles at any simple roots of $4\gamma^2[\varepsilon - V(x)] + [V'(x)]^2$, but the associated residues can be shown to be zero. It is also clear that none of those poles can lie on the path of integration. Further, the same observations apply for $\langle \mathcal{K}_1 \rangle$ and $\langle \mathcal{K}_2 \rangle$.

Finally, note that the asymptotic estimates for the moments given here treat the eigenvalue $\varepsilon$ as a variable that can take any value. That is, their evaluation does not require the choice of the correct quantized values. It is then important that the correct eigenvalues are obtained, be it by the SAFE-based approach described in Section 2.3 or by other techniques[16,17], in order to use them in the moment expansions.

*3.2 Specific potentials for demonstration*

For demonstration, the Pöschl-Teller potential gives a simple test case that now permits an investigation of the accuracy of Eq.(51) for determining the rms widths of the associated bound states. For this case, it follows from symmetry that $\mathcal{M}_1 = 0$ and it turns out that $\langle X \rangle$, $\langle \mathcal{K}_0 \rangle$, and $\langle \mathcal{K}_1 \rangle$ are all zero, and it is easy to evaluate $\langle \mathcal{K}_2 \rangle$ numerically. The relative error in the asymptotic approximation of the rms width is plotted in Fig.4 for the bound states of Fig.3. Because the bound states are not known in closed form, these errors are evaluated by reference to moments of numerically integrated solutions. Notice that, even at zeroth order, none of the errors exceed 10%. Further, the second-order correction gives a striking improvement, e.g. the relative errors in the rms widths are then typically below $10^{-4}$.

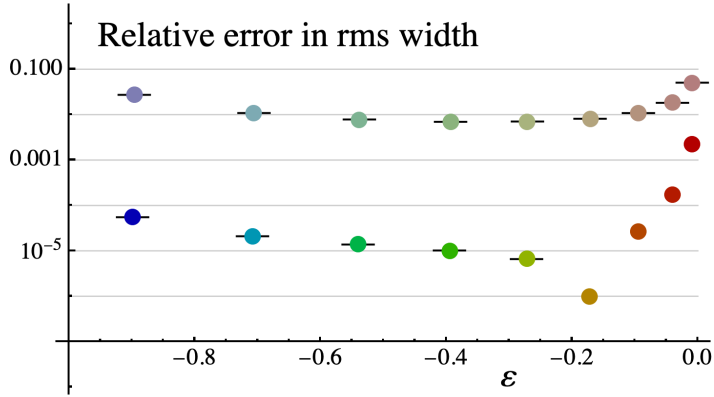

**Figure 4** Plot of the relative error in estimates of $\sqrt{\mathcal{M}_2/\mathcal{M}_0}$ for the bound states in Fig.3. The upper (dimmed) points correspond to the familiar semi-classical term in Eq.(51) while the lower points include the wave-related correction. The colouring follows the convention of Fig.3 and points where the error is negative carry a short horizontal line in order to clarify that a sign change is associated with the anomalously low error of one of the points.

Crucially, the evaluation of these asymptotic estimates for the mean square widths is far simpler than the alternative, especially for larger $n$. With growing $n$, the field needs to be sampled increasingly densely to perform the quadrature to compute the moments via alternative means. What's more, non-trivial computation is required to approximate each required field value with sufficient accuracy. The evaluation of the asymptotic estimate, on the other hand, is relatively trivial because $\langle \mathcal{K}_2 \rangle$ is given by a single integral whose integrand is relatively slowly varying. Further, the results become more accurate with increasing $k$ and are entirely independent of $\gamma$ and depend only on $\varepsilon$.

A reassuring check on the validity of these asymptotic results follows upon investigating how their errors decay as the asymptotic limit is approached. For example, a log-log plot of the error in Eq.(51) can give a valuable confirmation that the terms up to and including those of $O(k^{-2})$ are all correct. Rather than working with the various eigenvalues for a specific value of $k$ as done for Fig.4, this is achieved by fixing $\varepsilon$ and investigating the error seen for the associated discrete values of $k$ that admit bound states. Upon solving Eq.(37) for $k$, it is therefore appropriate to adopt

$$k = k_n = \frac{\sqrt{(2n+1)^2-(1+\varepsilon)}+(2n+1)\sqrt{-\varepsilon}}{2(1+\varepsilon)} = \frac{2n+1}{2(1-\sqrt{-\varepsilon})} - \frac{1}{4(2n+1)} + O\left(\frac{1}{(2n+1)^3}\right), \tag{55}$$

and —for a variety of values of $\varepsilon$— compare the results given by Eq.(51) with the mean square widths of the numerically determined bound states for $k = k_n$ as $n$ increases. Such a data set is presented in Fig.5. Note that, while the exact expression for $k_n$ in Eq.(55) must be used when numerically integrating to determine the bound states and their widths, it is only the asymptotic approximation shown there that is adopted when evaluating Eq.(51). That asymptotic approximation can also be derived directly from Eq.(42).

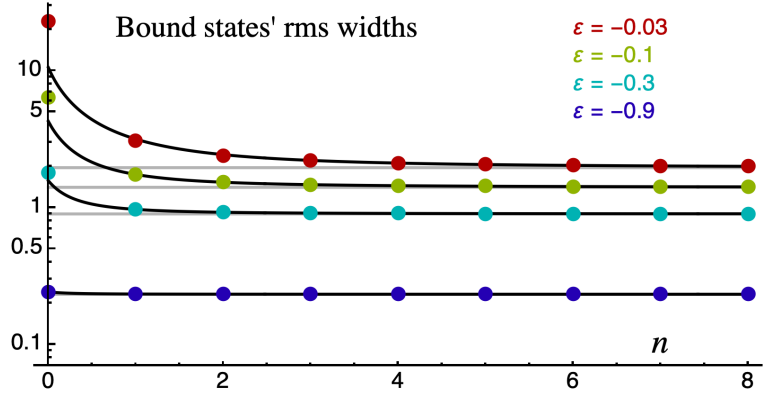

**Figure 5** A log plot of $\sqrt{\mathcal{M}_2/\mathcal{M}_0}$ for the lowest nine bound states at four fixed values of $\varepsilon$, i.e. when $k = k_n$ of Eq.(55) for $n = 0,1,2,..8$. The gray horizontal lines correspond to the zeroth-order approximation of Eq.(51) —a constant for any given energy level— whereas the wave-related correction of second order is included for the much more accurate black curves.

It can be seen in Fig.5 that, for bound states at energies near the bottom of the well, i.e. $\varepsilon$ near $-1$, the rms width is nearly constant and in close agreement with the semi-classical approximation. At energy levels that are close to escaping the well, however, the ground state (see the red point with $n = 0$) can have an rms width that is more than an order of magnitude larger than that given by the basic semi-classical approximation $\langle X^2 \rangle$. As shown by the black curves, the second-order correction in Eq.(51) significantly boosts accuracy. Notice that the coloured points in Fig.5 are defined only for the numerically determined discrete bound states, while the asymptotic approximation of Eq.(51) is obviously a readily computed continuous function of $k$, hence also of $n$ by way of Eq.(55).

A more interesting view of the same data is presented in Fig.6 and reveals a surprising property: If our analysis had not extended beyond zeroth order, it would have been reasonable to expect the errors presented in the upper curves in Fig.6 to exhibit errors of $O(k^{-1})$, hence have slope $-1$. Keep in mind that, for fixed $\varepsilon$, it follows from Eq.(55) that $n = O(k)$. The actual error is evidently of $O(k^{-2})$, as also indicated by Eq.(51). When its second-order term is retained, on the other hand, Eq.(51) suggests that the error is expected to be of $O(k^{-3})$ but the results presented in Fig.6 reveal that the next non-zero correction is of $O(k^{-4})$. This helps to explain the striking accuracy delivered by that asymptotic result.

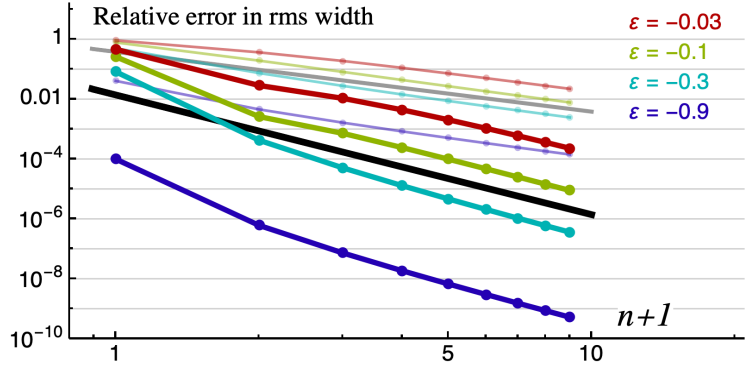

**Figure 6** A log-log plot of the relative error in $\sqrt{\mathcal{M}_2/\mathcal{M}_0}$ for four fixed values of $\varepsilon$ when $k = k_n$ of Eq.(55) for $n = 0,1,2,..8$. The upper (dimmed) points are for the zeroth-order approximation of Eq.(51) whereas the wave-related correction of second order is included for the lower points. Note that the horizontal axis is taken to be $n + 1$ (rather than just $n$) so that $n = 0$ can be included. The gray and black sloping lines have gradient $-2$ and $-4$, respectively.

Similar results are found for the Morse potential, but because the bound states are known in closed form for that case, that example is only of value for code checking. Other potentials that have received a lot of attention are polynomials. The most well known of course is the simple harmonic oscillator for which $V(x) = x^2$. This provides a basic testing ground where SAFE yields the exact bound states for a particular value of $\gamma$ [namely the one that makes $\bar{Q}(x)$ constant] and, as they must, the wave-related corrections are found to vanish in Eq.(51) because the semi-classical result is exact for this particularly simple case. More generally, monomial potentials lend themselves to the application of operator methods for determining moments[18]. Even so, the quartic potential is used here as a further demonstration, namely

$$V(x) = x^4. \tag{56}$$

Neither the eigenvalues nor the eigenfunctions can be found in closed form for the potential in Eq.(56). Matrix methods have been reported for accurately calculating the eigenvalues along with tabulated results[17]. A plot of the associated low-order bound states is presented in Fig.7. With a simple homogeneous potential like that in Eq.(56), it is readily seen that changing $k$ in Eq.(1) serves only to rescale the eigenvalues and the independent coordinate; the accuracy of the asymptotic approximation is then found to improve for increasingly higher eigenstates. That is, the asymptotic parameter $k$ can effectively be replaced by the eigenstate index $n$. This stands in contrast to the Morse and Pöschl-Teller potentials where there is a finite number of bound states and where $k$ and $n$ play independent roles.

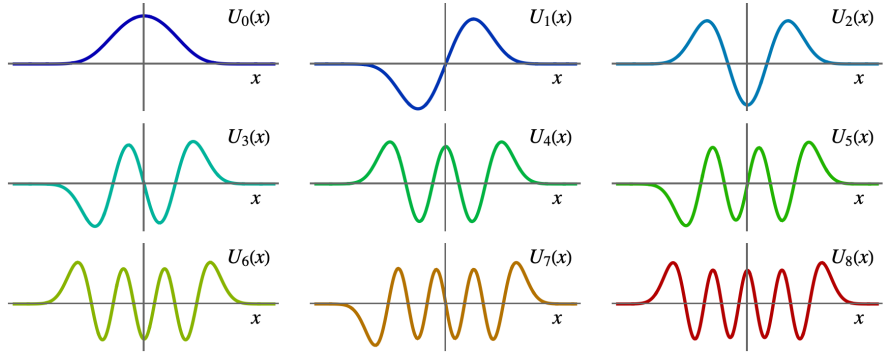

**Figure 7** Plots of the first nine bound states for the quartic potential of Eq.(56) found by numerically integrating Eq.(1). Unlike those in Fig.3, all plots are on the same horizontal scale and changing the value of $k$ simply rescales those axes equally. Because the axes are not given explicit scales, these plots present the general case.

The classical motions for the quartic potential are given in terms of Jacobi elliptic functions. It turns out that, by using Eqs.(30)-(33), $A_1(\tau,\gamma)$ and $A_2(\tau,\gamma)$ can be found in closed form and the analogue of Eq.(42) for this case is

$$\varepsilon_n^{(1)} = \left\{ \frac{\sqrt{2\pi}3\beta(2n+1)}{4k} \left[ 1 + \sqrt{1 + \frac{4}{3\pi(2n+1)^2}} \right] \right\}^{4/3}, \tag{57}$$

where $\beta = \pi/[\Gamma(1/4)]^2$. Because the scaling arguments reveal that the exact eigenvalues are also proportional to $k^{-4/3}$, the relative error in the asymptotic approximation is a function of $n$ alone and the plots in Fig.8 are valid for all $k$. It turns out that, for this example, the elements of Eq.(51) can be evaluated in closed form to find:

$$\frac{M_2}{M_0} = 8\,\beta^2\sqrt{\varepsilon} + \frac{1-576\beta^4}{48\,\varepsilon\,k^2} + O(k^{-3}). \tag{58}$$

To better appreciate this result, it is appropriate to solve Eq.(57) for $k$ to find the analogue of Eq.(55):

$$k_n = \frac{(2n+1)\beta}{\sqrt{2\pi}}\,\varepsilon^{-\frac{3}{4}}\left[3\pi + \frac{1}{(2n+1)^2} + O\left(\frac{1}{(2n+1)^3}\right)\right]. \tag{59}$$

With this, Eq.(58) becomes

$$\frac{M_2}{M_0} = \sqrt{\varepsilon}\left[8\,\beta^2 - \frac{576\,\beta^2 - \beta^{-2}}{216\,\pi(2n+1)^2} + O\left(\frac{1}{(2n+1)^3}\right)\right]. \tag{60}$$

Because the scaling arguments reveal that the exact eigenvalues are similarly proportional to $\sqrt{\varepsilon}$, it follows that the errors in the asymptotically estimated rms widths are functions of only $n$. That is, unlike the situation in Fig.6, there is now only one plot for all cases. The associated results presented in Fig.8 are derived by using numerically determined bound states. Note that it can be seen that the error in Eq.(60) is actually of $O((2n+1)^{-4})$. It is striking that the asymptotic results yield impressive accuracy without going to extreme cases, e.g. the error in the rms width is only about ten parts per million for $n = 4$ and less than one part per million for $n$ as low as 9 since it decreases proportionally to $n^{-4}$.

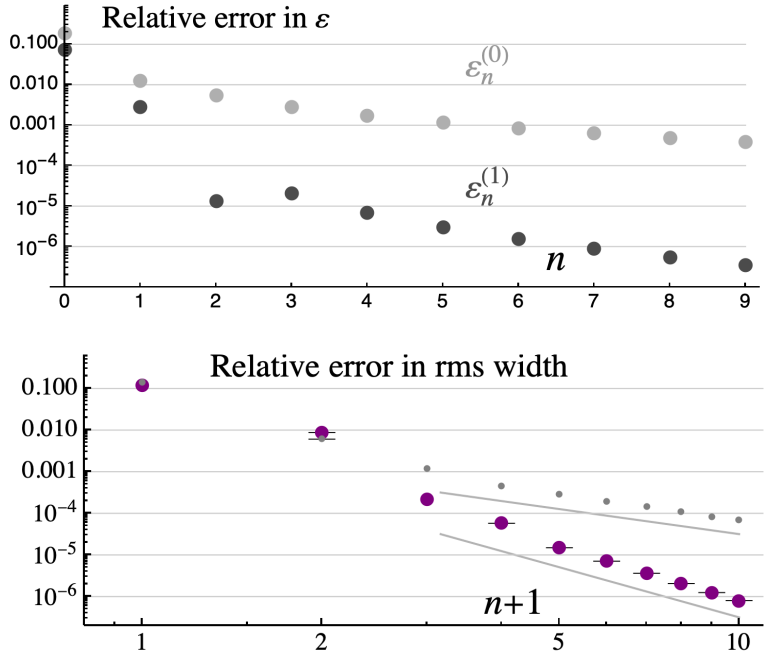

**Figure 8** Plots analogous to Figs.1 and 6, but now for the quartic potential of Eq.(56) and using Eqs.(57) and (60). The sloping lines once again have gradient $-2$ and $-4$.

## 4. Concluding Remarks

The asymptotic methods applied in Sec. 3 yield estimates of the moments of bound states such as those given in Eqs.(49) and (51) that effectively bypass evaluation of the associated wavefields. Taken together with the asymptotic expansions for the eigenvalues —as developed within the SAFE framework in the paragraph following Eq.(34)— the results have been shown to give surprisingly accurate approximations. Notice that following the change of variables introduced in Eq.(53), each of the key components, namely $\langle \mathcal{K}_0 \rangle$, $\langle \mathcal{K}_1 \rangle$, and $\langle \mathcal{K}_2 \rangle$, is given as a definite integral involving just the potential, i.e. $V(x)$, and its derivatives. Remarkably, the associated integrands involve SAFE's width parameter, namely $\gamma$, but the associated definite integrals were shown to be precisely independent of $\gamma$. Curiously, however, we have been unable to rework those integrals to explicitly eliminate $\gamma$. The fact that the integrals are ill-defined for either $\gamma = 0$ or $\gamma = \infty$ suggests to us that an approach based on localized field contributions such as SAFE may be essential to the derivation of these results. We know of no alternative option for deriving these asymptotic expansions.

It may also be worth observing that, as pointed out in Eqs.(I-6.4) and (II-3.1), once a field is expressed in the form given in Eq.(23), its Fourier transform has an analogous form. It follows that momentum moments can be asymptotically approximated following an essentially identical process to that described above.

One of the surprises that emerged in this work is the fact that, although the SAFE bound state estimate is given as an expansion with both even and odd powers of $k^{-1}$, see Eqs.(23) and (28), there appears to be no odd powers present in the resulting expansions for the moments. As a step towards an explanation, we noted that the conventional asymptotic solution can be expressed in a form where the series for the field amplitude involves only even powers, see Eqs.(15) and (16). While the associated singularities at turning points obscure the connection with field moments, one of SAFE's strengths is that it characteristically avoids those singularities while working from nothing more than the real-valued classical trajectories.

Finally, we comment that our primary objective for SAFE was to construct better asymptotic field estimates based on just the zeroth-order coefficient in the expansion for $A(\tau, \gamma)$ while exploiting its asymptotic first-order correction (or even just an approximation for it) to arrive at accuracy estimates. The results reported here, however, required that the second-order term also be included in order to determine a non-zero wave-based correction for these moments. That is, the novel result in Eq.(32) was essential for this effort. Curiously, the error in the end results turns out to be of $O(k^{-4})$ and this partially explains the accuracy of Eqs.(49) and (51).

## Acknowledgements

MAA acknowledges funding from the Excellence Initiative of Aix Marseille University –A*MIDEX, a French 'Investissements d'Avenir' programme. The authors also thank Ron Gordon for useful discussions regarding integration.